\documentclass[twocolumn,showpacs,aps,pre,
groupedaddress,amssymb,amsmath,nobalancelastpage]{revtex4}

\usepackage{graphicx}
\usepackage{longtable}

\begin{document}

\title{Dislocation Dynamics in an Anisotropic Stripe Pattern}

\author{Carina Kamaga}
\author{Fatima Ibrahim}
\author{Michael Dennin}

\affiliation{Department of Physics and Astronomy, University of
California, Irvine. Irvine, CA 92696-4575}

\date{\today}

\begin{abstract}

The dynamics of dislocations confined to grain boundaries in a
striped system are studied using electroconvection in the nematic
liquid crystal N4. In electroconvection, a striped pattern of
convection rolls forms for sufficiently high driving voltages. We
consider the case of a rapid change in the voltage that takes the
system from a uniform state to a state consisting of striped
domains with two different wavevectors. The domains are separated
by domain walls along one axis and a grain boundary of
dislocations in the perpendicular direction. The pattern evolves
through dislocation motion parallel to the domain walls. We report
on features of the dislocation dynamics. The kinetics of the
domain motion are quantified using three measures: dislocation
density, average domain wall length, and the total domain wall
length per area. All three quantities exhibit behavior consistent
with power law evolution in time, with the defect density decaying
as $t^{-1/3}$, the average domain wall length growing as
$t^{1/3}$, and the total domain wall length decaying as
$t^{-1/5}$. The two different exponents are indicative of the
anisotropic growth of domains in the system.

\end{abstract}

\pacs{47.54.+r, 64.60.Cn}

\maketitle

\section{Introduction}

The dynamics of topological defects are observed to dominate the
temporal evolution of patterns in many physical systems. However,
our understanding of the quantitative contribution of the defect
dynamics to the evolution of patterns is still not complete. One
area in which topological defects potentially play a central role
is the growth of domains in a patterned system after a sudden
change in the external parameters. For many striped systems, such
as diblock copolymers and convection in fluids, many of the
topological defects (such as disclinations and domain walls) are
the same as those that exist in classic models for the growth of
``uniform domains'', such as Ising models or x-y models
\cite{CH93}. For uniform systems, the contribution of the
topological defects to the time evolution of the domains is
relatively well understood \cite{REV}; whereas, this is not the
case for striped systems.

It is useful to briefly review the situation for uniform domain
growth \cite{REV}. One is generally interested in the evolution of
a system after a rapid change of an external parameter, or a
quench. Typically, one considers an initially uniform state that
immediately after the quench is no longer an equilibrium or
steady-state phase of the system. The new equilibrium phase is
degenerate, and domains of the different states form. For example,
in an Ising system, one would have domains of up and down spins.
During the subsequent evolution of the system, or {\it
coarsening}, the domains are characterized by a single length
scale. This length grows as a power of time, $t^n$, where $n$ is
the growth exponent. We use the designation ``uniform domains'' to
refer to systems in which within a domain, the system is uniform.
Examples of this type of system include magnetic systems, metallic
alloys, binary fluids, and nematic liquid crystals \cite{REV}. One
of the main goals of this field is to understand the possible
values of $n$ and other features of the late time scaling. For
uniform systems, much of the late-time behavior can be understood
in terms of the topological defects of the order parameter
\cite{REV}. For most cases, the growth exponent is $1/3$ if the
order parameter is conserved, such as in binary fluids, and the
exponent is $1/2$ if it is not conserved, as in magnetic systems
\cite{REV}. Various interesting alternate cases exist, especially
for the x-y model in one and two dimensions \cite{REV}. In
contrast, no similar general framework exists for patterned
domains. These are systems in which within a domain, the system
exhibits a pattern, such as stripes.

Stripes, or more generally patterns, occur in a wide range of
systems \cite{CH93,GL99}, including convecting fluids, animal
coats, polymer melts, and ferromagnets. Stripes occur both as an
equilibrium state of the system, such as in diblock copolymers,
and as a result of external driving, as in convection in fluids. A
sudden change of an external parameter, a {\it quench}, can bring
the system from a spatially uniform state to a striped state that
undergoes coarsening, or phase ordering. The coarsening of striped
domains has focused on {\it isotropic} systems
\cite{EVG92,CM95,CB98,HAC00,BV01,HCSH02}, with a significant
fraction of the work focusing on simulations of model equations,
such as the Swift-Hohenberg equation
\cite{EVG92,CM95,CB98,BV01,QM02}. Experimental work has been
limited to two systems, one isotropic (diblock copolymers
\cite{HAC00,HCSH02}) and one anisotropic (electroconvection in
nematic liquid crystals \cite{PD01,KFD04}). For isotropic systems,
the stripes can have any orientation. In this case, the dominant
defects are disclinations and domain walls, though dislocations
are observed as well. For diblock copolymers, the measured growth
exponent is consistent with $n = 1/4$. Detailed studies of the
disclination dynamics are able to explain this measured exponent
\cite{HAC00,HCSH02}.

For simulations of isotropic systems, growth exponents are usually
consistent with $1/4$ \cite{EVG92,CM95,CB98} or $1/5$
\cite{EVG92,CM95,CB98,PD01}. Factors affecting the measured
exponent include external noise and the quantity used to
characterize the domains. There is evidence that for small enough
quench depths the growth exponent is $1/3$ \cite{BV01,BV02,QM02}.
An open question is the connection between the simulations and the
experiments with the diblock copolymer systems. The dominant
defects appear to be different between the simulations and the
experiment, leaving open the question of a general explanation for
the coarsening behavior.

For electroconvection (an anisotropic system), two main classes of
patterns occur as the initial transition: normal and oblique rolls
\cite{BZK88,KP95,RWTSS89}. Electroconvection uses a nematic liquid
crystal. The molecules of a liquid crystal align on average along
a particular axis, referred to as the director \cite{GP93}. Normal
rolls consist of a system of parallel rolls oriented with the
wavevector parallel to the director field (the average axis of
alignment for the molecules). The main defect in the normal roll
regime is isolated dislocations. The dynamics of these
dislocations have been studied both experimentally and
theoretically \cite{GPRS89,RSR90,KBP90,KBPTZ89}, though not in the
context of domain growth. Oblique rolls correspond to stripes in
which the wavevector forms a nonzero angle with respect to the
undistorted director field. In the oblique roll regime, a quench
typically produces a pattern consisting of domains of stripes with
only two orientations, referred to as zig and zag rolls. In this
case, the main defects are domain walls and dislocations.  For a
particular case of electroconvection, growth exponents of $1/4$
were observed \cite{PD01}. However, in contrast to isotropic
systems, disclinations were not present in this system. Again,
this points to the need to better understand the dynamics of the
various classes of defects if a general framework for
understanding phase ordering in pattern forming systems is to
emerge.

In considering the existing work, there are at least two obvious
questions. Given that the uniform and pattern forming systems
exhibit similar topological defects, why are the resulting growth
exponents so different? Given the range of growth exponents
observed for pattern forming systems, can they even be explained
in terms of the dynamics of topological defects? An important step
in answering both of these question is the elucidation of the
defect dynamics. The work with diblock copolymers has already made
important contributions along these lines for disclinations. In
this paper, we focus on the dynamics of confined dislocations
found in the oblique roll regime of electroconvection
\cite{KBPTZ89}.

Figure~\ref{singleimage}b is a top view of a pattern in an
electroconvection cell that illustrates the defects of interest in
this paper. The cell consists of a nematic liquid crystal confined
between specially treated glass plates that align the director
parallel to the plates along a single axis. This axis is defined
to be the x-axis, or horizontal direction. The y-axis, or vertical
direction, is also parallel to the glass plates, but perpendicular
to the undistorted director. The z-axis is taken perpendicular to
the plates. The nematic liquid crystal is doped with ionic
impurities. An ac voltage is applied perpendicular to the plates.
There exists a critical value of the applied voltage, $V_c$, at
which a transition from a spatially uniform state to a striped
state occurs. The striped state consists of convection rolls with
a corresponding periodic variation of the director and charge
density. As shown in Fig.~\ref{singleimage}, we studied the case
of {\it oblique rolls}, and the two classes of topological defects
are domain walls and dislocations. The dislocations are special in
that they are mostly confined to vertical domain walls.
Disclinations do not occur, as the stripes are not easily curved.
As we will show, the domain walls are essentially static, so only
the dynamics of the dislocations are of interest. In this paper,
we report on qualitative features of the dislocation dynamics and
their interactions. In addition, we will report on the time
dependance of three global measures of defect kinetics:
dislocation density, average domain wall length, and the total
domain wall length per area. The rest of the paper is organized as
follows. Section II describes the experimental details. Section
III presents the results, and Sec. IV is a summary and discussion
of the results.

\section{Experimental Details}

The details of the experimental apparatus are described in
Ref.~\cite{D00}. The nematic liquid crystal N4 was doped with 0.1
wt\% of tetra n-butylammonium bromide [${\rm(C_4H_9)4N Br}$].
Commercial cells \cite{EHCO} with a quoted thickness of 23~$\mu$m
and 1~cm$^2$ electrodes were used, giving an aspect ratio of 435.
The average wavelength of the rolls was $51\ {\rm \mu m}$. The
sample temperature was maintained at $30.0 \pm 0.002\ {\rm
^{\circ}C}$. The patterns were observed from above using a
modified shadowgraph setup \cite{RHWR89,ASR99,D00} that emphasized
the contrast between zig and zag rolls. The magnification was
chosen to monitor the largest possible area of the sample while
maintaining enough resolution to resolve the stripe pattern. The
area imaged contained approximately 150 rolls.

All of the results reported here are for a fixed quench depth of
$\epsilon = V^2/V^2_c - 1 = 0.05$.  After a quench, domains of zig
and zag rolls form within 60 seconds. (The relevant relaxation
times for electroconvection are the director relaxation time
($0.6\ {\rm s}$) and the charge relaxation time ($4 \times
10^{-5}\ {\rm s}$).) The domains are separated by two classes of
grain boundaries. There are domain walls that extend horizontally.
Across these grain boundaries, the phase of the stripes is
continuous. The vertical grain boundaries consist of discrete
steps that are formed by dislocations. Each step contains either
an extra zig or zag roll. Both of the grain boundaries are
illustrated in Fig.~\ref{singleimage}, schematically in
Fig.~\ref{singleimage}a and using an actual image of the system in
Fig.~\ref{singleimage}b. Figure~\ref{singleimage}a illustrates the
Burgers vector construction for one of the dislocations,
demonstrating the extra $2\pi$ of phase that occurs when
traversing a closed path around the dislocation core. Vertical
boundaries on opposite sides of a domain are always composed of
dislocations of opposite sign. Also shown is the definition of the
domain wall length, $L_i$, used later in the analysis of
coarsening.

The dislocation number density, $\rho(t) = n(t)/A$, was measured
by counting the number of dislocations ($n(t)$) in a fixed viewing
area $A$. Because of the resolution used to take the images, and
the fact that the defects were in such close proximity to each
other within the domain walls, the defects were counted by hand.
At each time, $\rho(t)$ was averaged over ten different quenches.
For the same times, we measured the length of the individual
horizontal domain walls in the system, $L_i$ (see
Fig.~\ref{singleimage}a). From this, we computed both the average
horizontal wall length ($<L> = \frac{1}{N}\sum L_i$, where $N$ is
the number of horizontal walls) and the total domain wall length
per area ($L = \frac{1}{A}\sum L_i$). Both of these quantities
were averaged over seven quenches. The results are plotted in
Fig.~\ref{time}. Here time is scaled by the director relaxation
time, $\tau_d = 0.6\ {\rm s}$ \cite{BZK88}. In order to avoid
issues of the initial growth of the amplitude and wavenumber, we
only considered times greater than $400 \tau_d$. After this time,
the average wavenumber changed by less than $0.3\%$.

\section{Results}

The first obvious difference between this system and previous
experiments \cite{HAC00,PD01} is the dramatic anisotropy of the
growth. The dislocations in the vertical grain boundaries move
essentially horizontally. Some discrete steps in the vertical
direction are observed. This is very different from the usual
glide or climb of a dislocations. This is best seen in a movie of
the motion. An archived movie made from snapshots taken every 60
seconds and with a playback rate of 0.3 s/frame can be found at
Ref.~\cite{movie}. This horizontal motion is directly responsible
for changes in the horizontal length scale of the domains.

In contrast to the dislocations, the horizontal grain boundaries
are effectively stationary. Any vertical ``motion'' of domain
boundaries occurs when an individual horizontal grain boundary is
eliminated. This elimination occurs when the oppositely charged
dislocations that form the domain wall's endpoints(see
Fig.~\ref{singleimage}) annihilate. If only a single pair of
endpoints annihilate, one has a discrete change of the vertical
length by 49 microns. This step size is set by the step height
corresponding to one dislocation in a vertically orientated domain
wall. In order to eliminate a domain completely, the entire set of
oppositely changed dislocations that form the vertical walls must
annihilate each other. This is illustrated by the series of
snapshots in Fig.~\ref{annihilation} taken from the archived movie
\cite{movie}. In this case, the number of dislocations on each
side of the domain differ by one. After the elimination of the
domain, a single dislocation remains. Many domains contained equal
numbers of dislocations, producing no isolated dislocations.

There are a small number of isolated dislocations, as seen in
Fig.~\ref{annihilation}. These dislocations move through the
sample either by climbing or gliding. Isolated dislocation either
eventually move into a domain wall or annihilate with an
oppositely charged dislocation in a domain wall.
Figure.~\ref{isolated} is a series of snapshots taken from the
archived movie \cite{movie} that illustrate the incorporation of
an isolated dislocation into a domain wall. This is consistent
with the behavior of isolated dislocation predicted in
Ref.~\cite{KBPTZ89}. They rarely annihilate with other isolated
dislocation because of their extremely low density.

During the evolution process, the dislocations often exhibit
behavior substantially more complex than that shown in
Fig.~\ref{annihilation}. For example, Fig.~\ref{osc}a-d
illustrates a case in which a single dislocation in a domain wall
is observed to be undergoing oscillatory motion. Eventually, this
defect is close enough to an oppositely charged defect that it
accelerates toward that defect and is annihilated (Fig.
~\ref{osc}e - h). Figure~\ref{bounce} illustrates another
interesting behavior. (This event is can also be found in the
archived movie \cite{movie}.) Entire walls of oppositely charged
defects can approach within some distance, and then move apart. In
a simple picture, oppositely charged defects would attract each
other and such a bounce would not be possible. However, these
defects are not isolated, and one must account for the entire
field of defects to describe these more complicated motions. This
is outside the scope of this paper and will be the subject of
future work. One also observes initially coherent walls of
dislocations undergoing dispersion as they move (Fig.~\ref{disp}).
This separation of dislocations occurs because the details of the
interactions result in dislocations within a wall moving with
different speeds. Finally, an extremely rare event is the
nucleation of a new zig or zag domain. What is interesting is that
extremely thin domains are possible in which the horizontal walls
are slightly curved and the ends to contain dislocations (see
Fig.~\ref{nucleate}).

Given the complex nature of the dislocation dynamics, for the
purposes of this paper, quantitative measures focused on the time
dependence of global quantities. The three quantities of interest
are the dislocation density, the domain wall density, and the
total domain wall length. The time dependence of these quantities
is able to provide information about the rate of ordering along
the axes parallel and perpendicular to the director. The results
are summarized in Fig.~\ref{time}.

One observes that all of the quantities are consistent with
power-law growth after $t \approx 1000 \tau_d$. Fits of the data
in this regime give: $\rho \sim t^{-0.32 \pm 0.02}$; $L \sim
t^{-0.18 \pm 0.07}$; and $<L> \sim t^{0.33 \pm 0.02}$. With the
limited range that we are able to observe, it is difficult to show
conclusively that a scaling regime has been reached. However, two
conclusions are clear: (a) the growth is anisotropic and is best
described by at least two different exponents; and (b) the
exponents are different from previously studied striped systems
\cite{HAC00,PD01}. The first point is illustrated in
Fig.~\ref{time}c, where the dashed line has a slope of $-1/3$. The
second point is illustrated by the dashed curves in
Fig.~\ref{time}a and b. Here, the dashed curves represent the
previously observed growth exponent of $1/4$.

Given that the defect motion is essentially confined to the
horizontal (or x) direction, we tested for the existence of two
different growth exponents. Typically, for domain growth, the
scaling hypothesis assumes all lengths are scaled by a single
scale factor $R(t) \sim t^n$, and area scales as $R^2$. For our
system, because of the obvious anisotropy, we assume that
horizontal lengths and vertical lengths of domains are scaled
independently by scale factors $L_x \sim t^{n_x}$ and $L_y \sim
t^{n_y}$, respectively. Therefore, the number of domains, $c$, in
our fixed viewing area $A$ scales as $c \sim A/ (L_x L_y)$.
Because the dislocations are confined in vertical walls, the
number of defects is given by the number of domains times a
typical vertical dimension of the domain: $n(t) \sim c L_y$. This
gives $\rho (t) = n(t)/A \sim L_y/(L_x L_y) \sim t^{-n_x}$. The
average horizontal domain wall length ($<L>$) scales as $<L> \sim
L_x \sim t^{n_x}$. As with the number of defects, the total
horizontal domain wall length scales as $c L_x$. Therefore, the
total domain wall length per viewing area will scale as $L \sim c
L_x/A = L_x/(L_x L_y) \sim t^{-n_y}$. The results in
Fig.~\ref{time} are consistent with two growth exponents, $n_x =
1/3$ and $n_y = 1/5$, with the growth in the vertical direction
being substantially slower than the growth in the horizontal
direction. Here, the time dependence of $\rho(t)$ and $<L>$
provide independent measures of $n_x$.

Another issue is the impact of the finite viewing window, which
determines the maximum time for which we can view the system. This
is seen in Fig.~\ref{time}c, where the possible scaling of $L(t)$
breaks down for $t > 5000 \tau_d$, even though the other measures
still exhibit possible scaling. This is understandable because
this is the time when most horizontal boundaries extend across our
field of view, and yet contain significant numbers of step
dislocations of opposite sign. Therefore, as these dislocations
annihilate with each other, both $\rho(t)$ and $<L>$ continue to
evolve, while $L(t)$ remains effectively constant.

\section{Summary}

We report observations of the dynamics of an interesting class of
dislocations: dislocations confined to vertically oriented domain
walls between two degenerate oblique rolls.  The dislocations
exhibit highly non-trivial behavior, forming coherent grain
boundaries, exhibiting motion that is neither climb nor glide, and
executing interesting dynamics, including oscillations. The motion
of isolated dislocations toward the horizontal grain boundaries
and the confinement of the dislocations to grain boundaries is
expected on general arguments from amplitude equations
\cite{KBPTZ89}. However, existing theoretical work focused on the
horizontal domain walls and individual dislocations
\cite{KBPTZ89}. A more detailed theoretical and experimental study
is required to understand the full range of observed behavior. For
example, it is clear that the observed dynamics are often due to
the interactions of many dislocations.

Measures of the global properties of the topological defects
provide some insight into the phase ordering of this system. On
the one hand, the combination of anisotropy and stripes results in
a relatively simple system. There are only two classes of defects
(domain walls and dislocations), and their basic motions are
straightforward. The dislocations move horizontally, and the
domain walls are essentially stationary. This suggests that the
phase ordering should be relatively straightforward to understand.
On the other hand, the system is an interesting example of how
stripes can make the system more complicated than the standard
uniform systems. This is best seen by comparing the system to two
standard universality classes for phase ordering in uniform
systems: Ising and X-Y models.

Aspects of the electroconvection patterns are analogous to an
Ising system, i.e. a systems of spins with two states. In our
system, the two states are the zig and zag rolls, and the
horizontal domain walls are the topological defects one would
expect in an Ising system. However, the presence of the stripes
results in additional topological defects: dislocations. These
defects are analogous to vortices in an x-y model, i.e. spins with
any orientation in the plane, in that both vortices and
dislocations have the same topological charge. In our system, they
occur predominantly in vertical domain walls or as steps between
two horizontal domain walls. Even with this ambiguity, because
there is no obvious conservation law (both zig rolls and zag rolls
are eliminated), one expects a growth exponent of $1/2$ for both
the Ising and X-Y model \cite{REV}. The observed growth for
electroconvection is clearly slower, being consistent with $1/3$
in one direction and $1/5$ in the other. This suggests that the
confined dislocations represent a new type of coarsening dynamics.

One direction that may provide further insight into the exact
cause of the slower dynamics is comparisons with simulations of an
anisotropic Swift-Hohenberg model \cite{B03}. Initial simulations
are consistent with a growth exponent of 1/3 for shallow quenches
with pinning effects becoming important as a function of quench
depth \cite{B03}. This will be the subject of future work.

\begin{acknowledgments}

The authors wish to thank Ben Vollmayr-Lee, Denis Boyer, Eberhard
Bodenschatz, and Werner Pesch for useful conversations. This work
was supported by NSF grant DMR-9975479. M. Dennin also thanks the
Research Corporation Cottrell Scholar and Sloan Fellowship for
additional funding for this work.

\end{acknowledgments}


\clearpage

\begin{figure}
\caption{\label{singleimage}(a) A schematic drawing illustrating a
transition from a zig to a zag and back to a zig region of the
system. The black lines represent lines of constant phase between
the stripes. There are two horizontal domain walls, one at each
transition. Also, one example of each type of step dislocation is
shown. The Burgers vector is also illustrated for one of the
dislocations. The length of a horizontal domain wall is indicated
by $L_i$. (b) A close up of a section of the system that shows two
step dislocations (black squares). The white squares are provided
as an aid in counting the zig and zag rolls. One finds that there
are two more zig rolls than zag, as expected with two
dislocations. The dashed line highlights one of the horizontal
domain walls. The scale bar represents 0.15 mm.}
\end{figure}

\begin{figure}
\caption{\label{annihilation}Four images separated by 60 seconds
illustrating the collapse of a zig domain. The scale bar
represents 0.15 mm.}
\end{figure}

\begin{figure}
\caption{\label{isolated} Images separated by 150 seconds
illustrating the incorporation of an isolated dislocation into a
domain wall. The scale bar represents 0.15 mm.}
\end{figure}

\begin{figure}
\caption{\label{osc}Eight images separated by 300 seconds
illustrating the motion of a single dislocation highlighted by a
circle in (a). A dust particle on the surface of the sample is not
removed to provide a frame of reference. In (a) - (d), the
dislocation oscillates. In (e) - (h), the dislocation is attracted
to one of opposite sign and annihilated. The scale bar represents
0.15 mm.}
\end{figure}

\begin{figure}
\caption{\label{bounce}
Images separated by 150 seconds illustrating oppositely charge
dislocations moving together, ``bouncing'', and ultimately
attracting each other. One can also observe the spread of
dislocations in a vertical wall that is highlighted in
Fig.~\ref{disp}. The scale bar represents 0.15 mm.}
\end{figure}

\begin{figure}
\caption{\label{disp} Images separated by 300 seconds illustrating
dislocations in a domain wall spreading out in time. The scale bar
represents 0.15 mm.}
\end{figure}

\begin{figure}
\caption{\label{nucleate} Images separated by 150 seconds
illustrating the nucleation of narrow domain without a dislocation
pair. The scale bar represents 0.15 mm.}
\end{figure}

\begin{figure}
\caption{\label{time}(a) Plot of $\log_{10}(\rho)$ versus
$\log_{10}(t)$. Symbols are experimental data. The solid line has
a slope $-0.32$. For comparison, the dashed line has a slope of
$-0.25$. (b) Plot of $\log_{10}(<L>)$ versus $\log_{10}(t)$.
Symbols are experimental data. The solid line has a slope $0.33$.
For comparison, the dashed line has a slope of $0.25$. (c) Plot of
$\log_{10}(L)$ versus $\log_{10}(t)$. Symbols are experimental
data. The solid straight line has slope of -0.2. For comparison,
the dashed line in (c) has a slope of -0.33.}
\end{figure}

\end{document}